\documentclass[journal,twoside,web]{ieeecolor2} 
\usepackage{generic}
\usepackage{cite}
\usepackage{amsmath,amssymb,amsfonts}
\usepackage{graphicx}
\usepackage{textcomp}
\makeatletter
\let\NAT@parse\undefined
\makeatother
\usepackage[hidelinks]{hyperref}
\begin{document}

\title{Characterizing Datasets to Anticipate Unsupervised Domain Adaptation Performance for Radiographic Pneumonia Detection}

\author{\mbox{Iftakhar~Hossain$^{a,*}$},~\mbox{Md.~Imran~Hasan$^{a,*}$},~\mbox{Salahuddin~Ahmed$^{c}$},~\mbox{Eric~D.~McCollum$^{b}$},\\\mbox{Abdullah~H.~Baqui$^{d}$},~and~\mbox{Taufiq~Hasan$^{a,e,\dagger}$},~\IEEEmembership{Senior~Member,~IEEE}%
\thanks{Manuscript received XXXX XX, 2026.}
\thanks{$^{*}$Iftakhar~Hossain and Md.~Imran~Hasan contributed equally
to this work and share first authorship.}
\thanks{$^{\dagger}$Corresponding author: Taufiq~Hasan.}
\thanks{$^{a}$Iftakhar~Hossain, Md.~Imran~Hasan and Taufiq~Hasan are with the
Department of Biomedical Engineering and the mHealth Laboratory, Bangladesh
University of Engineering and Technology (BUET), Dhaka-1205, Bangladesh
(e-mail: \mbox{iftakhar.bme.buet20@gmail.com}; \mbox{imran220.284@gmail.com};
\mbox{taufiq@bme.buet.ac.bd}).}
\thanks{$^{b}$Eric~D.~McCollum is with the Global Program in Pediatric Respiratory
Sciences, Eudowood Division of Pediatric Respiratory Disease, Johns Hopkins
University School of Medicine, Baltimore, MD, USA.}
\thanks{$^{c}$Salahuddin~Ahmed is with the Projahnmo Research Foundation,
Dhaka, Bangladesh.}
\thanks{$^{d}$Abdullah~H.~Baqui is with the International Center for Maternal and
Newborn Health, Department of International Health, Johns Hopkins Bloomberg 
School of Public Health, Baltimore, MD, USA.}
\thanks{$^{e}$Taufiq~Hasan also has a secondary affiliation with the Center for
Bioengineering Innovation and Design, Johns Hopkins University, Baltimore, MD, USA.}}

\maketitle
\begin{abstract}
\textit{Objective:} Deep learning models for pneumonia detection from chest
radiographs exhibit degraded accuracy when applied across institutions, which 
motivates Unsupervised Domain Adaptation (UDA). We observe that UDA performance depends on the properties of both the training and test datasets, and the performance gains are asymmetric with respect to dataset pairs. 
%Adaptation helps for some pairs of datasets and not others,
%and the same dataset pair can provide different results depending on which one is used for training. 
%However, training an adaptation model for every candidate pairing is costly in
%the low-resource settings where the problem matters most. 
In this study, we investigate which structural properties of the X-ray datasets are most important predictors of UDA performance, and whether they can be measured before adaptation is attempted.
\textit{Methods:} Five chest radiograph datasets were characterized by the
class separability of their features, obtained from one forward pass of a
frozen DenseNet-121 and summarized by the Silhouette Score, with the Fisher
Ratio as a supporting measure. All twenty directed source-target pairs were
adapted with a Domain-Adversarial Neural Network (DANN) while direct transfer without
adaptation was also evaluated. Next, seven published transferability
estimators were computed on the same representation for comparison, and
their significance was assessed for predicting UDA performance. 
%by exact permutation tests at a unit of analysis matched to each question.
\textit{Results:} We found that adaptation performance depended on both endpoints in opposite
directions. More separable target datasets attained higher AUC scores
($r=0.934$, $p=0.042$), whereas more separable sources produced lower 
AUC scores ($r=-0.922$, $p=0.017$). The signed difference in separability tracked the
signed difference in AUC between the two directions of a pair
($r=-0.932$, $p=0.017$) and identified the better direction in 10 of 10 pairs
($p=0.001$). On evaluation of a target dataset, the proposed descriptor was comparable to the best published
estimators. %,  none of them addresses the source or the direction.
\textit{Conclusion:} Source and target datasets seem to play the opposite structural roles in adaptation. Thus, a dataset with higher class separability is easier as a test set, but not as good for training a model for adaptation.
%, which a target-centric account cannot express.
\textit{Significance:} In case of pneumonia detection from X-ray images, the proposed framework can be used to decide which dataset is better for adaptation towards a target in a low-resource setting without running full experiments. 
%Although the experiments here are based on pneumonia detection from X-ray images, we believe the findings can generalize towards other application domains. 
%With five datasets these findings are exploratory and require confirmation at larger scale. Within that scope, separability computed in advance
%provides a low-cost means of anticipating the direction in which adaptation between two chest radiograph
%domains is likely to succeed, identifying the better direction in all
%ten pairs examined at a cost of five forward passes rather than twenty adaptation runs.
\end{abstract}


%%%%%%%%%%%%%%%%%%%%%%%%%%%%%%%%%%%%%%%%%%%%%%%%%%%%%%%%%%%%%%%%%%%%%%%%%%%%%%%
\section{Introduction}
\label{sec:intro}

Pneumonia is the single largest infectious cause of death in children worldwide,
killing 740,180 children under the age of five in 2019 and accounting for 14\% of all deaths
in this age group~\cite{b37}. Chest radiography is the established reference standard
for identifying the alveolar consolidation that marks a pneumonia episode, and it is inexpensive,
widely deployable, and well suited to case confirmation in field and epidemiological studies~\cite{b18}.
However, a radiograph becomes diagnostic only once it has been read, and reliable 
interpretation requires trained radiologists or pediatricians working to a standardized protocol,
a resource which is both scarce and geographically mal-distributed across the low-income settings
where the pneumonia burden is concentrated~\cite{b17,b18}. Thus, automated pneumonia detection 
from chest radiographs, deployed at the point of care alongside the portable radiography units
already in use in such settings, can play a vital role in early case identification
where expert readers are not available. 

Deep learning models for this task can reach
expert-level accuracy when trained and tested on a single well-curated dataset~\cite{b1}.
This accuracy is not always retained when the model is deployed at another institution. 
The cause is domain shift: differences in imaging equipment, acquisition protocol,
patient population and disease prevalence move the distribution of the new data away from the training
data~\cite{b2}. The resulting degradation is a barrier to safe
cross-institutional deployment, and it is most severe in the settings where such models would provide the greatest benefit.

In low-resource settings, two additional constraints apply. Locally labeled radiographs are scarce, 
since expert reading is the same resource that is in short supply, and the computational budget for 
retraining is limited. A common strategy is thus to utilize a labeled public dataset from a
different population as the source and to bridge the gap with Unsupervised Domain
Adaptation (UDA)~\cite{b27,b28,b29,b30}. Training a separate adaptation model for
each directed source-target pair is expensive, which makes it valuable to know
in advance which pairings are likely to adapt well.

Adaptation is also directional. Adapting from domain A to domain B need not
perform as well as adapting from B to A, a phenomenon known as
\emph{asymmetric transfer}~\cite{b3}. A useful characterization of transfer
should thus explain not only whether adaptation succeeds but why its success depends
on which domain supplies the labels. Existing transferability estimators, reviewed
in Section~\ref{sec:relatedwork}, describe transfer principally through the target domain
and treat the source as a fixed feature extractor.
However, these estimators do not consider the fact that the source domain's own class structure
may contribute independently, and to the best of our knowledge no published measure characterizes
the source in a way from which the preferred direction of transfer between two domains could be anticipated.

In this study, we characterize chest radiograph domains by the class structure of their
features, measured before any adaptation model is trained, and relate that characterization
to the transfer performance observed afterwards. The aim is to establish which structural
properties of the source and of the target accompany successful adaptation, and thus to
anticipate whether, and in which direction, transfer between two domains is likely to succeed. 
The study is confined to pneumonia detection from chest radiographs; whether the
same relationships hold for other pathologies, tasks or modalities is untested.
With five datasets the analysis is exploratory and is reported as such. The
main contributions are as follows.

\begin{itemize}
\item We propose a pre-computable structural description of a chest radiograph
dataset, a single bounded scalar from one pass of a frozen feature extractor,
and relate it to adaptation performance over five datasets and twenty directed
pairs.

\item We confirm that stronger target class separability accompanies better
adaptation, consistent with the cluster assumption, and benchmark the
descriptor against seven published transferability estimators on the same
representation.

\item We observe that source class structure is informative in the opposite
direction: more overlapped sources transfer better. No published estimator
describes the source this way.

\item We show that transfer asymmetry follows the signed difference in
separability, which identifies the better direction in every pair under
adaptation. A target-only score cannot express this: one value per domain
predicts the same outcome both ways.

\item We verify that these relationships hold in sign across both regimes, two
independently trained extractors and two separability measures, indicating
domain properties rather than artifacts of one algorithm, representation or
metric.
\end{itemize}

\begin{figure}[!t]
\centerline{\includegraphics[width=\columnwidth]{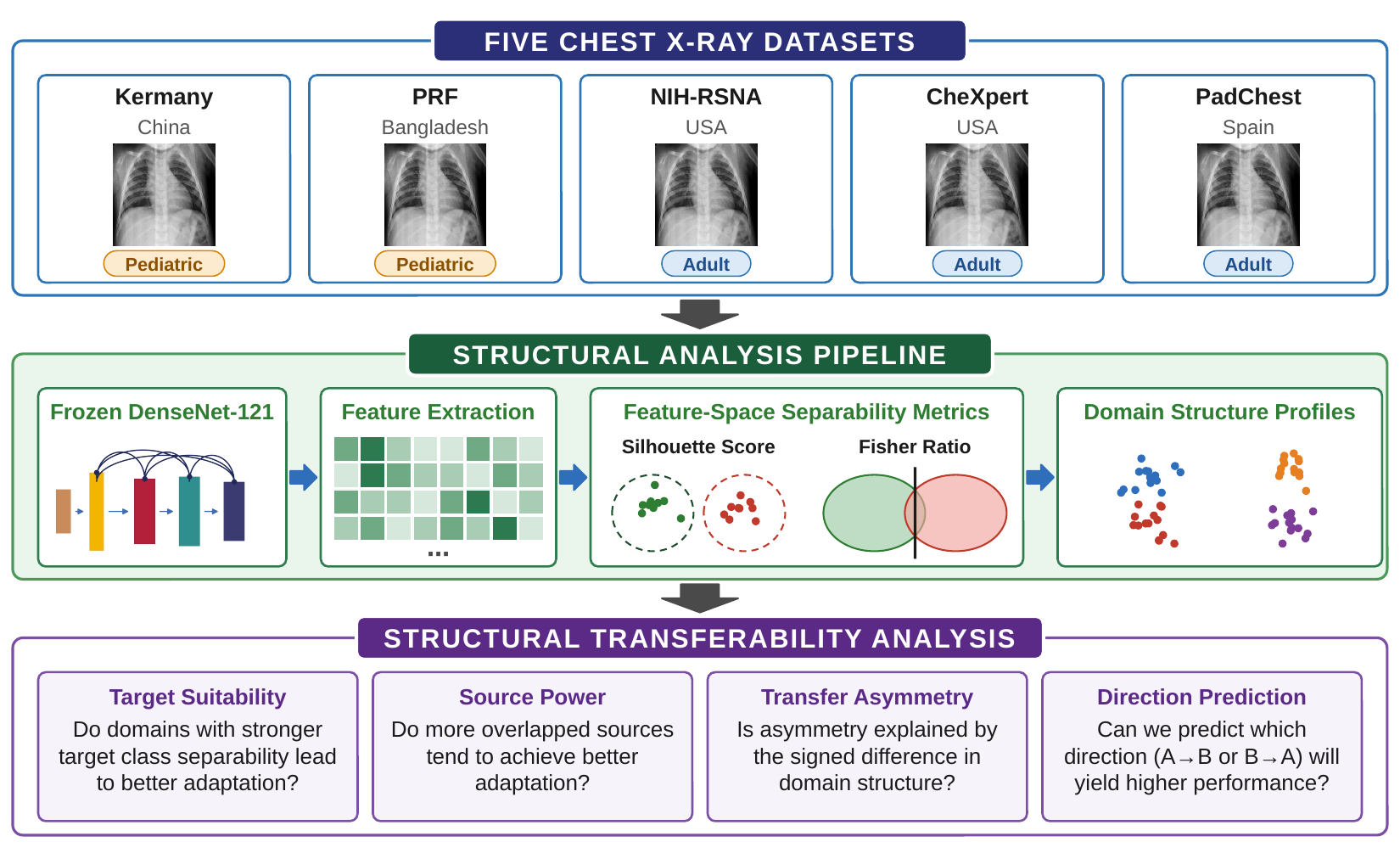}}
\caption{Overview of the structural analysis pipeline. Five chest radiograph
datasets are passed through a frozen DenseNet-121 to extract penultimate-layer
features, from which the class separability of the two diagnostic classes is
computed for each dataset, yielding a structural profile of each domain
independent of any particular source-target pairing.}
\label{fig:pipeline}
\end{figure}

%%%%%%%%%%%%%%%%%%%%%%%%%%%%%%%%%%%%%%%%%%%%%%%%%%%%%%%%%%%%%%%%%%%%%%%%%%%%%%%
\section{Related Work}
\label{sec:relatedwork}

\subsection{Transferability Estimation}
\label{sec:transferability}

Transferability estimation predicts how well a model will transfer without
running the full training procedure. The earliest practical measures score a
target dataset from a single forward pass of a pretrained model. H-score
\cite{b6} treats transferability as a ratio of between-class to total feature
covariance; LEEP \cite{b4} computes the log expected empirical prediction of
the target labels under the source classifier's output distribution; and LogME
\cite{b5} maximizes the evidence of a linear model fitted on the frozen
features. A later analysis showed that H-score is limited mainly by covariance
estimation in high dimensions and proposed a shrinkage estimator that recovers
much of its accuracy \cite{b32}.

Four measures introduced since 2021 refine this idea in different directions.
NLEEP \cite{b33} replaces the source classifier of LEEP with a Gaussian mixture
fitted to the target features, which removes the need for a source head
altogether. TransRate \cite{b34} estimates the mutual information between
features and labels through a coding-rate surrogate. PACTran \cite{b35} derives
a PAC-Bayesian bound on the fine-tuning risk and reports Gaussian and Dirichlet
variants. SFDA \cite{b36} embeds the features in a regularized Fisher space and
adds a self-challenging step that perturbs easy examples. GBC \cite{b7}
approaches the same problem through the geometry of the target classes: each
class is modeled as a Gaussian in the feature space of a pretrained network,
and the pairwise Bhattacharyya coefficient measures the overlap between the
class-conditional distributions, with lower overlap indicating higher expected
transferability. The argument is that this coefficient bounds the Bayes error
of the corresponding problem.

Although these measures differ in construction, they share a premise: that the
separability of the target classes under a given representation governs how
well a model transfers into that domain. The same principle arises
independently in adaptation theory, where the cluster assumption holds that
adaptation succeeds more readily when target features form discriminative
clusters through which a decision boundary can pass in a low-density
region~\cite{b8,b9}. Target-side separability as a predictor of transfer is
thus established, and we treat it here as a reference against which the
proposed descriptor is checked rather than as a finding; seven of these
estimators are computed on the present representation for that purpose
(Section~\ref{sec:charac}).

None of these measures provides a characterization of the source. In
each case the source enters only as a frozen feature extractor, and the score
is a property of the target alone. None treats the source domain's own class
structure as a separate factor, and thus none yields a quantity from
which the preferred direction of transfer between two domains could be
anticipated. A further limitation concerns evaluation rather than formulation.
These metrics were developed and validated primarily on natural-image
benchmarks, and their behavior in medical imaging, where the class boundary is
diagnostic rather than object-based and datasets are smaller and more
heterogeneous in acquisition, has been examined far less systematically.

\subsection{Domain Adaptation for Chest Radiographs}
\label{sec:cxrDA}

Domain adaptation for chest radiographs has been an active area. Early work
utilized supervised adaptation, in which a small amount of labeled target data is
available alongside a labeled source set. Feng \emph{et al.} proposed a deep
supervised domain adaptation network for pneumonia diagnosis that aligns the
source and target feature distributions while sharing the feature extraction
layers between a source-specific and a target-specific classifier~\cite{b10}.
Following a comparable strategy, supervised adversarial adaptation was applied
to the cross-population shift between chest radiograph datasets from different
countries, again relying on labeled target examples~\cite{b11}. That approach
builds on the Domain-Adversarial Neural Network (DANN) framework of Ganin
\emph{et al.}~\cite{b12}, which trains a shared feature extractor to be
simultaneously discriminative for the label-prediction task and agnostic to
domain origin.

A second family translates images between domains with generative adversarial
networks. CX-DaGAN utilizes an image-to-image translation network, guided by a
similarity-based source selection step, to synthesize source images resembling
the target domain for pneumonia diagnosis on a small target
dataset~\cite{b13}. C2ADA extends this with a cycle-consistent adversarial
module that aligns source and target features without paired images~\cite{b14}.
More recently, fully unsupervised approaches have removed the need for any
labeled or paired target data. TUNA-Net adapts adult chest radiographs to
resemble pediatric radiographs for pneumonia detection using only unlabeled
target images, recovering a substantial part of the performance lost to domain
shift~\cite{b15}.

Across this progression the source domain is treated either as fixed or as a
passive feature extractor, and the choice of source data is usually made
without reference to any measured property of the source itself. CX-DaGAN is a
partial exception in that it selects source \emph{images} by similarity to the
target, but it does not characterize the source domain as a whole. This holds
for most domain adaptation work in medical image analysis. The present study
addresses that gap by describing source and target on equal terms and relating
both descriptions to observed transfer performance.

%%%%%%%%%%%%%%%%%%%%%%%%%%%%%%%%%%%%%%%%%%%%%%%%%%%%%%%%%%%%%%%%%%%%%%%%%%%%%%%
\section{Methods}
\label{sec:methods}

\subsection{Datasets}
\label{sec:datasets}

Five chest radiograph datasets from different institutions, geographic regions
and patient populations are utilized. All are cast as binary classification into
\emph{Normal} and \emph{Pneumonia}; for datasets carrying additional disease
labels only the relevant classes are retained.
Table~\ref{tab:dataset_summary} summarizes the class distributions, with
datasets ordered as in every subsequent table by the separability reported in
Section~\ref{sec:charac}.

\textbf{Kermany} is a public pediatric collection acquired at a single
pediatric medical center during routine care, with every radiograph graded for
quality and diagnosis by expert physicians before release~\cite{b16}. It
contains 1,583 normal and 4,273 pneumonia images.

\textbf{NIH-RSNA} is built from the NIH ChestX-ray14 collection~\cite{b20},
whose pneumonia annotations were augmented by radiologists for the RSNA
Pneumonia Detection Challenge~\cite{b31}. A balanced subset of 3,350 normal and
3,350 pneumonia images was utilized.

\textbf{PadChest} is a large annotated collection from a Spanish healthcare
institution~\cite{b21}. A balanced subset of 3,350 normal and 3,350 pneumonia
images was extracted.

\textbf{CheXpert} is a large collection built from routine clinical imaging at
Stanford~\cite{b19}. Because the original release carries multiple thoracic
pathologies with uncertainty labels, a balanced binary subset of 3,350 normal
and 3,350 pneumonia images was constructed.

\textbf{PRF} is a private pediatric collection acquired at outpatient clinics
in the Projahnmo Research Foundation study area in the Sylhet district of rural
Bangladesh, as part of a pneumococcal conjugate vaccine effectiveness
study~\cite{b17}. Children aged 3--35 months with suspected pneumonia underwent
supine anteroposterior chest radiography on portable analogue units and images
were digitized using computed radiography cassette readers; radiographs were
interpreted by a panel of trained radiologists and pediatricians following the
standardized methodology for chest radiograph interpretation in
epidemiological studies~\cite{b18}. A subset was excluded during quality
control, leaving 2,777 normal and 1,999 pneumonia images.

\textbf{Splits and label handling.} Each dataset is partitioned once into a
training portion and a disjoint held-out evaluation portion, and that partition
is utilized for every role the dataset plays. The training portions contain 5,232
radiographs for Kermany (1,349 normal, 3,883 pneumonia), 6,000 for each of
NIH-RSNA, PadChest and CheXpert (3,000 per class), and 4,107 for PRF (2,388
normal, 1,719 pneumonia); Kermany follows the official split released with the
dataset. All structural descriptors of Section~\ref{sec:structural} are
computed on the training portion alone, so that a domain is characterized using
only data available before any adaptation model is trained.

When a dataset acts as the \emph{target}, its labels are discarded. The
adaptation procedure of Section~\ref{sec:regimes} sees the images of the target
training portion without labels; the target labels are utilized once only, on the
disjoint held-out portion, to compute the reported AUC. No label from the
held-out portion of any dataset enters training in either regime.

\begin{table}[!t]
\caption{Summary of the Datasets, Ordered by the Separability of
Table~\ref{tab:descriptors}}
\label{tab:dataset_summary}
\centering
\setlength{\tabcolsep}{4pt}
\begin{tabular}{lcccc}
\hline
\textbf{Dataset} & \textbf{Normal} & \textbf{Pneumonia} & \textbf{Total}
& \textbf{Train} \\
\hline
Kermany (pediatric)  & 1,583 & 4,273 & 5,856 & 5,232 \\
NIH-RSNA (adult)     & 3,350 & 3,350 & 6,700 & 6,000 \\
PadChest (adult)     & 3,350 & 3,350 & 6,700 & 6,000 \\
CheXpert (adult)     & 3,350 & 3,350 & 6,700 & 6,000 \\
PRF (pediatric)      & 2,777 & 1,999 & 4,776 & 4,107 \\
\hline
\end{tabular}
\end{table}

\subsection{Feature Representation}
\label{sec:feature_rep}

Motivated by the observation that pixel-space statistics entangle class-relevant structure
with acquisition nuisance such as scanner, exposure and positioning, structural descriptors
are computed not on raw pixels but on a fixed learned feature representation.
Let $D$ be one of the five datasets, consisting of $N$
grayscale radiographs $\{I_i\}_{i=1}^{N}$ with binary labels
$y_i \in \{0,1\}$. The primary extractor is a
DenseNet-121 trained for chest radiograph classification on the NIH
ChestX-ray14 collection, the dataset on which CheXNet~\cite{b1} was originally
trained; the weights are \texttt{densenet121-res224-nih} as distributed with
the TorchXRayVision library~\cite{b26}. The network defines a mapping
\begin{equation}
\boldsymbol{\phi}: \mathcal{I} \rightarrow \mathbb{R}^{1024},
\label{eq:phi}
\end{equation}
from image space $\mathcal{I}$ to the 1024-dimensional activation of the
penultimate layer, taken after the final rectification and global average
pooling. Each image is passed through the network once in inference mode with
no gradient updates, yielding $\mathbf{x}_i = \boldsymbol{\phi}(I_i)$. One
forward pass per dataset is required and no training, which is substantially
cheaper than training an adaptation model for each directed pair.

It should be noted that the extractor weights were trained on NIH ChestX-ray14,
which is the source of one of the five evaluated datasets. Section~\ref{sec:robust} accordingly repeats
the entire analysis with an independently trained DenseNet-121 (\texttt{densenet121-res224-chex},
trained on CheXpert) and reports the comparison.

The resulting vectors are standardized to zero mean and unit variance along
each of the 1024 dimensions, using statistics fit jointly over both classes of
that dataset. The overall pipeline is illustrated in Fig.~\ref{fig:pipeline}.

\subsection{Structural Descriptors}
\label{sec:structural}

The class structure of a dataset in this feature space is summarized by two
scalars. Let $\mathbf{X} = \{\mathbf{x}_i\}_{i=1}^{N} \subset \mathbb{R}^{d}$,
$d = 1024$, denote the set of standardized feature vectors of dataset $D$,
partitioned into the subsets $\mathbf{X}_0$ and $\mathbf{X}_1$ of the Normal
and Pneumonia classes with sizes $n_0$ and $n_1$. Throughout, bold lowercase
denotes a vector and plain italic a scalar.

\textbf{Silhouette Score.} A dataset has well-separated classes if its points
lie closer to others of their own class than to the opposite class. For a point
$\mathbf{x}_i$ of class $k$, let $a(i)$ be its mean Euclidean distance to the
remaining points of class $k$ and $b(i)$ its mean Euclidean distance to the
points of the opposite class. The silhouette coefficient of $\mathbf{x}_i$ is
\begin{equation}
s(i) = \frac{b(i) - a(i)}{\max\{a(i), b(i)\}},
\label{eq:silhouette_point}
\end{equation}
and the dataset-level Silhouette Score is its mean,
\begin{equation}
s(D) = \frac{1}{N}\sum_{i=1}^{N} s(i),
\label{eq:silhouette}
\end{equation}
with values closer to one indicating stronger separation~\cite{b22}. For
tractability $s(D)$ is estimated on a random subsample of the training portion
of up to 3,000 points with a fixed seed, since the underlying pairwise-distance
computation is $\mathcal{O}(N^2)$.

\textbf{Fisher Ratio.} As a supporting measure constructed on an entirely
different principle, the Fisher Ratio expresses separation as a ratio of
variances rather than of distances~\cite{b23}. A multivariate,
dimension-summed form is utilized. With $\mathbf{c}_0, \mathbf{c}_1 \in
\mathbb{R}^{d}$ the class centroids, the between-class term is the squared
Euclidean distance between them,
\begin{equation}
B(D) = \lVert \mathbf{c}_0 - \mathbf{c}_1 \rVert_2^{2},
\label{eq:fisher_between}
\end{equation}
and the within-class term is the sum of per-dimension unbiased variances over
both classes,
\begin{equation}
W(D) = \sum_{j=1}^{d} \mathrm{Var}(\mathbf{X}_0)_j
     + \sum_{j=1}^{d} \mathrm{Var}(\mathbf{X}_1)_j .
\label{eq:fisher_within}
\end{equation}
The Fisher Ratio is then
\begin{equation}
F(D) = \frac{B(D)}{W(D) + \epsilon},
\label{eq:fisher}
\end{equation}
with $\epsilon = 10^{-12}$ for numerical stability and larger values
indicating greater discriminability. This is a computationally efficient
multivariate generalization of the classical between/within variance ratio,
using summed per-dimension variances rather than a full $1024 \times 1024$
covariance matrix. It should not be confused with the scalar Fisher
discriminant of Linear Discriminant Analysis (LDA).

The Silhouette Score is adopted as the primary descriptor and the Fisher Ratio
is reported alongside it throughout. The Silhouette Score is bounded on
$[-1,1]$ and scale-free: it depends on neither feature magnitude,
dimensionality, nor sample count, which matters when comparing five datasets
that differ in size and acquisition characteristics. The Fisher Ratio is
unbounded and measures a different geometric quantity, so agreement between the
two is evidence that a finding is not an artifact of either construction. As
reported in Section~\ref{sec:charac}, the two agree completely on the ordering
of the five datasets, and we exploit that fact in
Section~\ref{sec:transferpotential}.

\subsection{Reference Transferability Estimators}
\label{sec:gbc}

To establish that the target-side relationship is a property of the domains, seven published estimators are
computed on the identical frozen representation and reported alongside the proposed descriptors:
H-score \cite{b6} and its shrinkage variant \cite{b32}, LogME \cite{b5},
NLEEP \cite{b33}, TransRate \cite{b34}, PACTran in its Gaussian form
\cite{b35} and SFDA \cite{b36}, together with GBC \cite{b7}. Every one of
these is a function of the features and the target labels alone, so all are
directly comparable to the proposed descriptors. LEEP \cite{b4} and the Dirichlet
variant of PACTran \cite{b35} are excluded because both require
$\theta(\mathbf{x}) = P(\text{source label} \mid \mathbf{x})$, a categorical
distribution over a source task's label set; the only head available here is
the pretrained multi-label pathology classifier, whose per-pathology sigmoid
outputs do not form such a distribution.

GBC is computed in the configuration recommended by its authors and is
described in full because it is the closest in spirit to the proposed descriptors. The
unstandardized vectors $\tilde{\mathbf{x}}_i = \boldsymbol{\phi}(I_i)$ are
projected onto their first $d' = 64$ principal components, which retain between
97.6\% and 98.4\% of the variance across the five datasets, and each class $k
\in \{0,1\}$ is modeled as a spherical Gaussian $\mathcal{N}(\boldsymbol{\mu}_k,
\sigma_k^{2} \mathbf{I}_{d'})$, where $\boldsymbol{\mu}_k$ is the class mean and
$\sigma_k^{2}$ the mean of the per-component unbiased variances. With $m =
(\sigma_0^{2} + \sigma_1^{2})/2$, the Bhattacharyya distance between the
class-conditional distributions is
\begin{equation}
D_B(0,1) = \frac{\lVert \boldsymbol{\mu}_0 - \boldsymbol{\mu}_1 \rVert_2^{2}}{8\,m}
+ \frac{d'}{2}\ln\frac{m}{\sigma_0 \sigma_1},
\label{eq:bhattacharyya}
\end{equation}
and the score of dataset $D$ is obtained by summing over ordered class pairs,
\begin{equation}
\mathrm{GBC}(D) = \sum_{k \neq l} D_B(k,l) = 2\,D_B(0,1),
\label{eq:gbc}
\end{equation}
larger values indicating greater separability. The score is reported in its
Bhattacharyya-distance form rather than as the exponentiated coefficient
$\mathrm{BC} = \exp(-D_B)$ utilized in \cite{b7}; the two are monotonically related
and give identical rankings, but across these five datasets the coefficient
spans roughly eight orders of magnitude, which makes linear comparison
uninformative. No reference estimator is ever combined with the proposed
descriptors, so agreement between them is an independent rather than a circular
check.

\subsection{Transfer Regimes}
\label{sec:regimes}

Every ordered pair among the five datasets is evaluated under two regimes.
Adaptation with DANN is the primary regime of this study; direct transfer
without adaptation is evaluated as a secondary regime, so that any
structure-performance relationship can be checked for consistency with and
without an adaptation procedure.

\begin{figure}[!t]
\centerline{\includegraphics[width=\columnwidth]{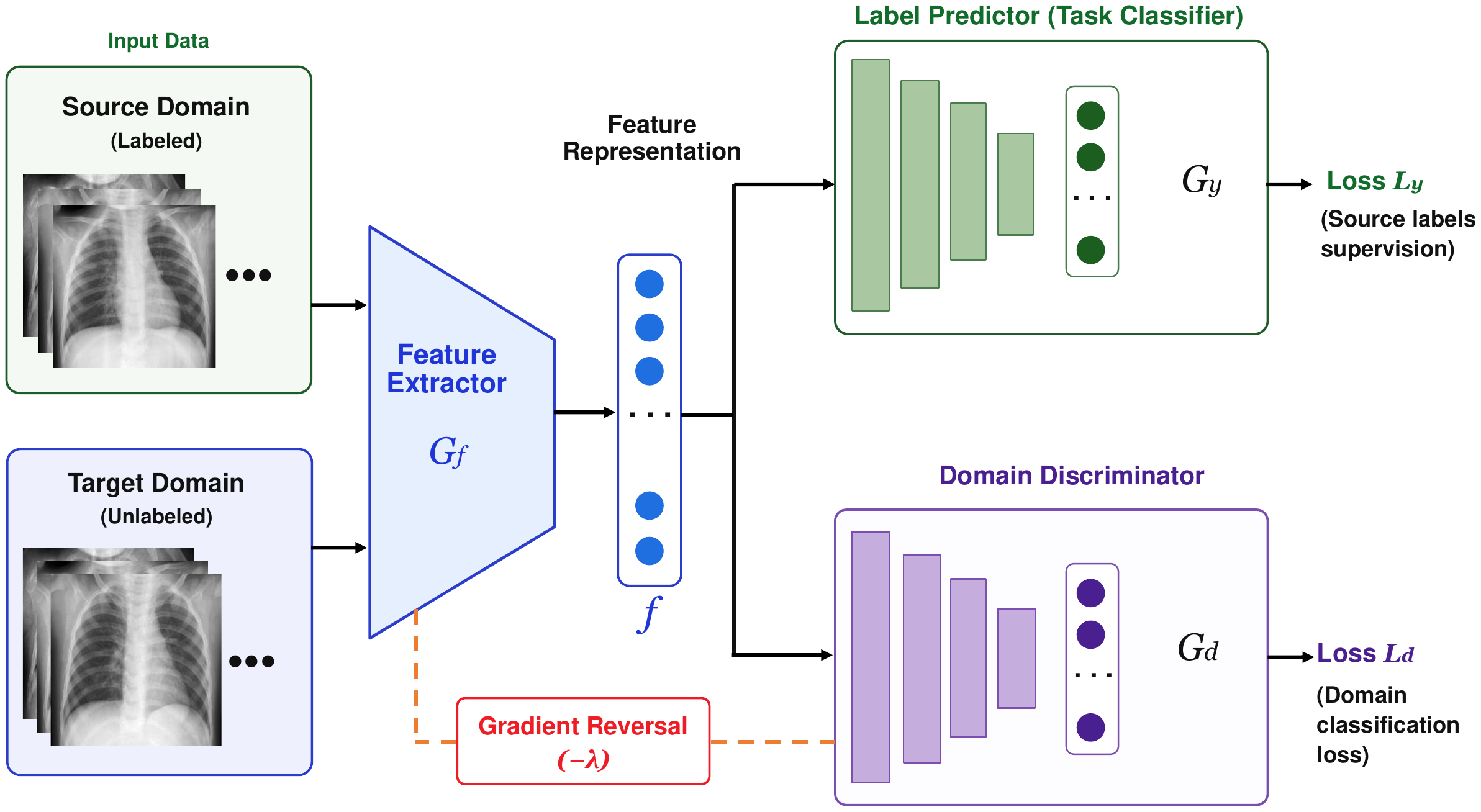}}
\caption{Domain-Adversarial Neural Network architecture. A shared feature
extractor $G_f$ maps labeled source and unlabeled target batches to a common
feature space $\mathbf{f}$. The label predictor $G_y$ is trained on source data
only (loss $\mathcal{L}_y$); the domain discriminator $G_d$ is trained on both
domains (loss $\mathcal{L}_d$). A gradient reversal layer between $G_f$ and
$G_d$ passes activations forward unchanged and negates the gradient during
backpropagation, driving $G_f$ toward domain-invariant features.}
\label{fig:dann_arch}
\end{figure}

\textbf{Adaptation with DANN.} The Domain-Adversarial Neural Network of Ganin
\emph{et al.}~\cite{b12} is utilized in its original form, without auxiliary losses
or architectural additions, so that an observed relationship can be attributed
to the domains rather than to the mechanics of a more elaborate method. The
architecture is shown in Fig.~\ref{fig:dann_arch}. Three components are trained
jointly: a shared feature extractor $G_f$, a label classifier $G_y$ and a
domain discriminator $G_d$, with parameters $\theta_f$, $\theta_y$ and
$\theta_d$. The feature extractor is driven to produce representations that are
discriminative for the task and uninformative of domain origin. Denoting the
label-prediction loss $\mathcal{L}_y$ and the domain-classification loss
$\mathcal{L}_d$, and given labeled source batches $\mathcal{D}_s$ and unlabeled
target batches $\mathcal{D}_t$ of sizes $n_s$ and $n_t$, the joint objective is
\begin{equation}
\mathcal{L}(\theta_f, \theta_y, \theta_d) =
\frac{1}{n_s}\sum_{i \in \mathcal{D}_s} \mathcal{L}_y^{(i)}
- \lambda \left( \frac{1}{n_s + n_t}
\sum_{i \in \mathcal{D}_s \cup \mathcal{D}_t} \mathcal{L}_d^{(i)} \right),
\label{eq:dann}
\end{equation}
where $\lambda > 0$ balances the two terms. The label classifier is trained on
source data only, since target labels are withheld
(Section~\ref{sec:datasets}); the domain discriminator sees both domains.
Optimizing $\theta_f$ to \emph{maximize} $\mathcal{L}_d$ directly would require
a second, opposing objective, so DANN instead inserts a gradient reversal layer
between $G_f$ and $G_d$: the identity on the forward pass, and a multiplication
of the gradient entering $G_f$ by $-\lambda$ on the backward pass. This allows
the network to be trained with one forward-backward pass and one optimizer
while $G_f$ is driven away from features that permit domain discrimination.
Following Ganin \emph{et al.}~\cite{b12}, $\lambda$ is annealed from 0 to 1 over training as
\begin{equation}
\lambda(p) = \frac{2}{1 + \exp(-10 p)} - 1,
\label{eq:lambda}
\end{equation}
where $p \in [0,1]$ is the fraction of total training iterations completed, so
that the domain term is suppressed while the label classifier is still poorly
conditioned. The full protocol is given in Table~\ref{tab:protocol}.

\textbf{Direct transfer.} A classifier is trained on a single source dataset
and applied to each target without modification, minimizing
\begin{equation}
\mathcal{L}_{\mathrm{direct}}(\theta_f, \theta_y) =
\frac{1}{n_s}\sum_{i \in \mathcal{D}_s} \mathcal{L}_y^{(i)},
\label{eq:direct_transfer}
\end{equation}
which is empirical risk minimization on the source alone with no reference to
the target during training. This regime represents deployment of a
source-trained model as-is and serves as a reference point for the effect of
domain shift without adaptation.

In both regimes the resulting classifier is evaluated on the held-out portion
of the target dataset by computing the area under the ROC curve between the
predicted pneumonia probability and the true label, yielding the directed AUC
reported for that pair. For each adapted pair the reported value is the highest
AUC attained on the target evaluation portion over the course of training. 
This value represents the best performance that adaptation attains for that pairing,
and it removes the stopping rule as a nuisance variable.
It is thus the appropriate quantity when the question is whether domain
structure governs what adaptation can achieve.
It is not an estimate of the performance a deployed model would obtain without
access to target labels.

\begin{table}[!t]
\caption{Training Protocol for the Adaptation Regime}
\label{tab:protocol}
\centering
\setlength{\tabcolsep}{4pt}
\begin{tabular}{ll}
\hline
\textbf{Component} & \textbf{Setting} \\
\hline
Feature extractor $G_f$    & ResNet-18, ImageNet-pretrained, \\
                           & final layer removed (512-d) \\
Label predictor $G_y$      & 512\,$\rightarrow$\,256, BN, ReLU, \\
                           & dropout 0.5, 256\,$\rightarrow$\,2 \\
Domain discriminator $G_d$ & identical head, 256\,$\rightarrow$\,2 \\
Gradient reversal          & $\lambda(p)$ per \eqref{eq:lambda} \\
\hline
Input size                 & $224 \times 224$, ImageNet normalization \\
Batch size                 & 32 per domain \\
Optimizer                  & Adam, weight decay $10^{-5}$ \\
Learning rate              & $10^{-4}$ \\
LR schedule                & reduce on plateau, factor 0.5, \\
                           & patience 3 \\
Gradient clipping          & max norm 5.0 \\
Epochs                     & 30, early stopping patience 7 \\
Class balancing (source)   & weighted random sampling \\
Random seed                & 42 \\
Reported AUC               & best value on the target evaluation \\
                           & portion over training \\
\hline
\end{tabular}
\end{table}

\subsection{Transfer Potential}
\label{sec:transferpotential}

To express the structural relationship between a source and a target as one
quantity, the Transfer Potential of a directed pair $(S \rightarrow T)$ is
defined as the difference in Silhouette Score between the two domains,
\begin{equation}
\mathrm{TP}(S \rightarrow T) = s(T) - s(S).
\label{eq:transferpotential}
\end{equation}
A positive value indicates a pairing in which the target is more separable than
the source and a negative value the converse.

The index is built from the Silhouette Score alone rather than from both
descriptors. As Section~\ref{sec:charac} reports, the Fisher Ratio and the
Silhouette Score agree on the ordering of the five datasets exactly
($\rho = 1.00$) and are almost perfectly collinear ($r = 0.999$), so the Fisher
Ratio carries no information beyond the Silhouette Score in this pool and a
composite of the two would add no predictive content while requiring an
arbitrary weighting. Between two interchangeable measures we take the bounded,
scale-free one. Because \eqref{eq:transferpotential} is a difference of raw
Silhouette Scores, it depends only on the two domains involved and not on the
pool of datasets under comparison.

\subsection{Statistical Analysis}
\label{sec:stats}

Structural descriptors are related to transfer performance, measured as AUC,
using the Pearson correlation. Because the number of datasets is small, the
Spearman rank correlation is reported alongside every Pearson coefficient, as
it assumes no linear form and is less sensitive to any single dataset.

Each question is evaluated at its appropriate unit of analysis. Target and
source roles are tested across the five datasets using mean AUC over the four
partner datasets. Directional effects are tested across the ten unordered pairs
by correlating the signed structural difference with the signed difference in
AUC between directions. Treating all twenty ordered pairs as independent would
repeat each dataset four times and understate $p$-values, which this choice
avoids.

Because asymptotic $p$-values are unreliable at five observations, significance
is assessed by permutation. Dataset-level analyses use the exact permutation
distribution over all $5! = 120$ assignments for both coefficients. For
pair-level analyses, where pairs share datasets, the five structural values are
permuted across datasets with AUC values held fixed. The $p$-value is the
proportion of the 120 assignments reaching at least the observed correlation
magnitude, giving a two-sided floor of $1/120 \approx 0.008$.

No correction for multiplicity is applied, and the results should be read as
exploratory rather than confirmatory. This is a limitation of scale rather than
procedure. With five domains, a Bonferroni threshold across these relationships
would require $p < 0.002$, below the $0.008$ permutation floor, making corrected
significance unattainable at this sample size. The analysis is thus
hypothesis-generating, emphasizing the direction and consistency of effects
rather than statistical significance.

%%%%%%%%%%%%%%%%%%%%%%%%%%%%%%%%%%%%%%%%%%%%%%%%%%%%%%%%%%%%%%%%%%%%%%%%%%%%%%%
\section{Results}
\label{sec:results}

\subsection{Transfer Performance Under Both Regimes}
\label{sec:perf}

\begin{figure*}[!t]
\centerline{\includegraphics[width=\textwidth]{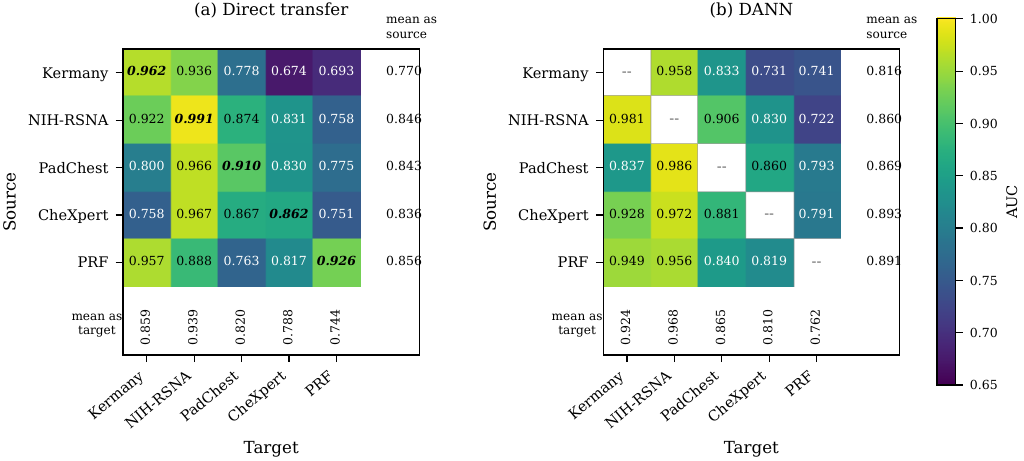}}

\caption{Directed AUC for all source-target pairs under (a) direct transfer
(source-trained model evaluated on the target without adaptation) and (b) DANN.
Rows are sources and columns are targets, both ordered by separability;
off-diagonal cells show the twenty cross-domain pairs. In (a), the diagonal
(bold italic) reports in-domain AUC, where a model is trained and evaluated on
the same dataset as a reference point rather than a transfer result. In (b),
the diagonal is undefined, since DANN requires distinct domains. The right
strip of each panel gives the mean AUC of each dataset as a source and the
bottom strip its mean AUC as a target, computed over cross-domain pairs only.
Both panels share a color scale.}

\label{fig:heatmap}
\end{figure*}

\begin{figure}[!t]
\centerline{\includegraphics[width=\columnwidth]{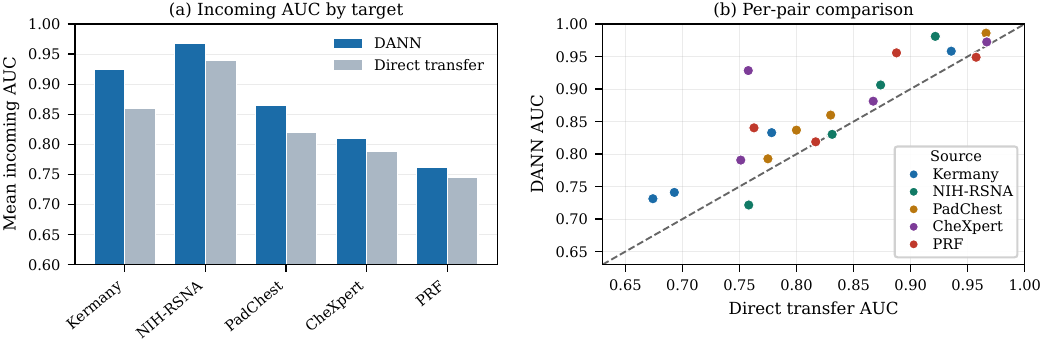}}
\caption{Transfer performance by regime. (a) Mean incoming AUC for each target
dataset under both regimes. (b) Per-pair AUC, points above the identity line
indicating pairs on which the adapted model reaches a higher value. Point
color denotes the source dataset.}
\label{fig:adaptation}
\end{figure}

Across the twenty ordered source-target pairs, adaptation with DANN attained a
mean AUC of 0.866 and direct transfer a mean AUC of 0.830, a relative improvement
of 4.3\% from adaptation. The complete set of directed AUC values under both regimes,
with the mean AUC of each dataset as a source and as a target, is shown in
Fig.~\ref{fig:heatmap}. Fig.~\ref{fig:adaptation} summarizes the per-target and
per-pair values.

From Fig.~\ref{fig:heatmap}, we observe that mean incoming AUC under adaptation
ranges from 0.762 for PRF to 0.968 for NIH-RSNA, a spread of 0.207, while mean outgoing AUC
ranges from 0.816 for Kermany to 0.893 for CheXpert, a spread of 0.077.
The target role thus accounts for the larger share of the variation and the source role for a
smaller but, as shown below, consistent one. The two extremes also invert
between the roles: Kermany is among the best targets and the worst source,
while PRF is the worst target and among the best sources.

\subsection{Structural Characterization of the Domains}
\label{sec:charac}

\begin{figure*}[!t]
\centerline{\includegraphics[width=\textwidth]{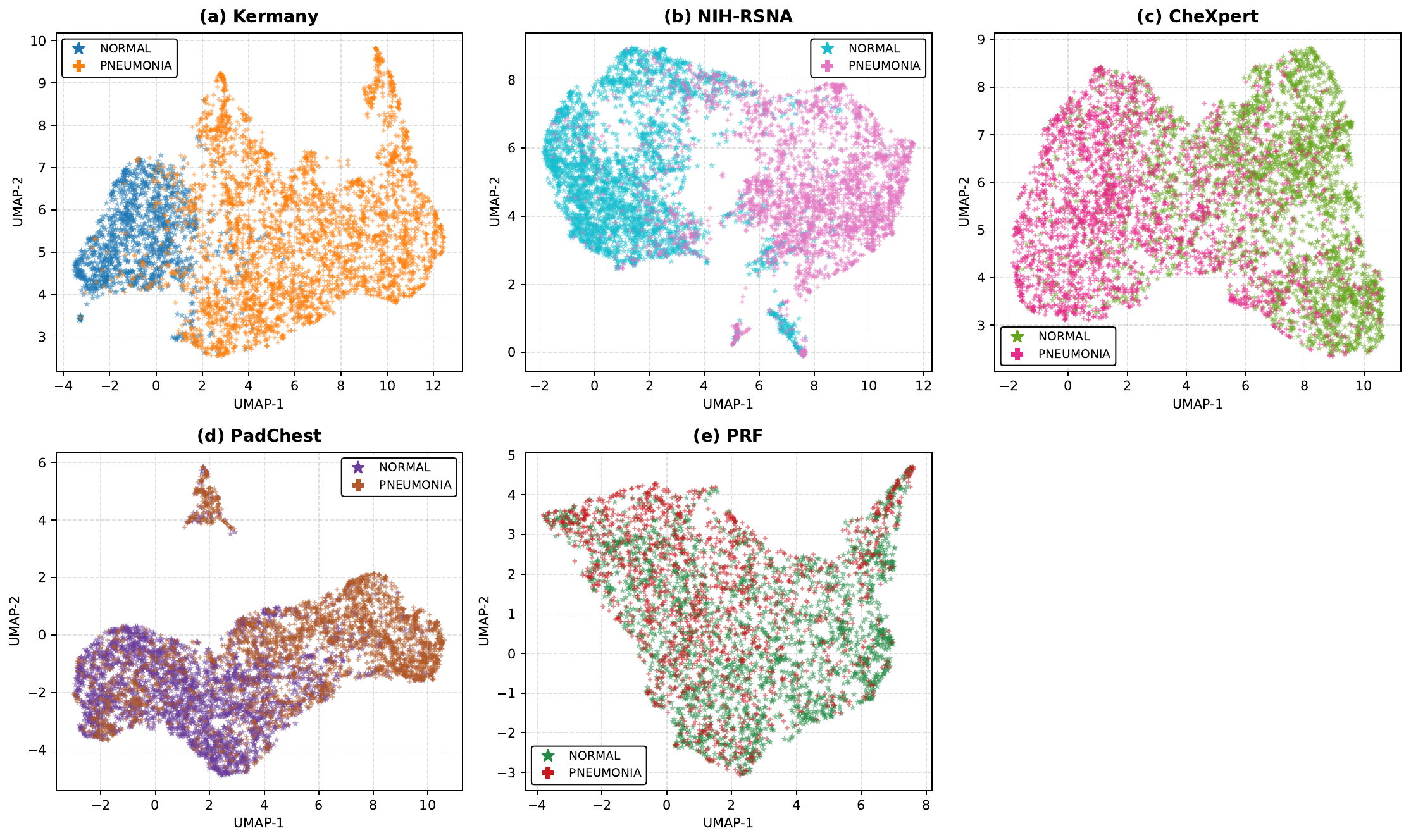}}
\caption{Two-dimensional UMAP embeddings of the DenseNet-121 feature
representation for each dataset, colored by class label. Datasets with a
higher Silhouette Score show more visibly separated classes.}
\label{fig:umap}
\end{figure*}

Table~\ref{tab:descriptors} reports the two proposed descriptors and the seven reference
estimators for each dataset on the 1024-dimensional representation. The
corresponding two-dimensional UMAP~\cite{b25} embeddings, colored by class,
appear in Fig.~\ref{fig:umap}.

From Table~\ref{tab:descriptors}, it is clear that all nine measures 
place the datasets in the same order: Kermany, NIH-RSNA,
PadChest, CheXpert, PRF. The agreement between the two proposed descriptors is
essentially total, with Spearman $\rho = 1.00$ and Pearson $r = 0.999$ between
the Silhouette Score and the Fisher Ratio, and $\rho = 1.00$ and $r = 0.975$
between the Silhouette Score and GBC. Separability in this representation is
thus not an artifact of any one construction: a distance-based measure, a
variance-based measure and a Gaussian-overlap measure agree exactly. This
redundancy justifies building the Transfer Potential of
\eqref{eq:transferpotential} on the Silhouette Score alone, as argued in
Section~\ref{sec:transferpotential}. The analyses below nonetheless report both
descriptors throughout, and no conclusion depends on which is utilized.

\subsection{Structure-Performance Relationships}
\label{sec:structperf}

The relationship between structure and transfer is examined from three
perspectives at three units of analysis: the role of the target alone, the role
of the source alone, and the signed difference between the two.

\textbf{The target dataset.} Table~\ref{tab:target} relates each dataset's separability
to its mean incoming AUC, together with the seven reference estimators computed
on the same representation. From this table, we observe that targets with better-separated 
classes receive higher AUC under adaptation ($r = 0.934$, $p = 0.042$ for the Silhouette Score;
$r = 0.932$, $p = 0.042$ for the Fisher Ratio). Every reference estimator shows
the same positive association, and the proposed descriptors sit among the
strongest: NLEEP is marginally higher at $r = 0.966$, the shrinkage H-score is
identical at $r = 0.934$, and the remainder fall between $0.840$ and $0.908$.
Nine of the ten measures, including both proposed descriptors, give $\rho = 0.900$,
differing from the AUC ordering by the single adjacent exchange of Kermany and
NIH-RSNA; the unshrunk H-score gives $\rho = 0.800$. The target-side
relationship is thus reproduced by every measure examined, which is the
expected outcome and the reason we treat it as a reference point rather than a
contribution.

\begin{table*}[!t]
\caption{Structural Descriptors and Reference Transferability Estimators per
Dataset, Primary Extractor. Higher Values Indicate Greater Class Separability
in Every Column. Datasets Are Ordered by Silhouette Score; All Nine Measures
Produce This Same Order.}
\label{tab:descriptors}
\centering
\setlength{\tabcolsep}{4.5pt}
\begin{tabular}{lcc|ccccccc}
\hline
& \multicolumn{2}{c|}{Proposed descriptors} & \multicolumn{7}{c}{Reference estimators} \\
Dataset & Silhouette & Fisher Ratio & GBC & LogME & H-score (s) & TransRate & NLEEP & PACTran-G & SFDA \\
\hline
Kermany   & 0.1535 & 0.3486 & 17.9147 & $+0.2347$ & 0.8637 & 60.6346 & $-0.3230$ & $-4.9686$  & $-0.2554$ \\
NIH-RSNA  & 0.1265 & 0.2891 & 11.3861 & $-0.0693$ & 0.7848 & 43.2202 & $-0.3432$ & $-4.3534$  & $-0.3702$ \\
PadChest  & 0.0671 & 0.1412 &  4.8135 & $-0.3791$ & 0.5890 & 30.5928 & $-0.5162$ & $-9.1413$  & $-0.4502$ \\
CheXpert  & 0.0326 & 0.0731 &  3.4402 & $-0.5465$ & 0.4330 & 20.1721 & $-0.6136$ & $-13.0551$ & $-0.5489$ \\
PRF       & 0.0086 & 0.0206 &  0.4197 & $-0.6024$ & 0.4024 & 14.8896 & $-0.6682$ & $-10.5807$ & $-0.5858$ \\
\hline
\end{tabular}
\end{table*}

\begin{table}[!t]
\caption{Target Structure and Reference Estimators versus Mean Incoming AUC
($n=5$; Exact Permutation $p$)}
\label{tab:target}
\centering
\setlength{\tabcolsep}{3pt}
\resizebox{\columnwidth}{!}{%
\begin{tabular}{lcccc|cccc}
\hline
& \multicolumn{4}{c|}{DANN} & \multicolumn{4}{c}{Direct transfer} \\
Measure & $r$ & $p$ & $\rho$ & $p_\rho$ & $r$ & $p$ & $\rho$ & $p_\rho$ \\
\hline
Silhouette Score        & $+0.934$ & 0.0417 & $+0.900$ & 0.0833 & $+0.850$ & 0.0583 & $+0.900$ & 0.0833 \\
Fisher Ratio            & $+0.932$ & 0.0417 & $+0.900$ & 0.0833 & $+0.852$ & 0.0583 & $+0.900$ & 0.0833 \\
\hline
GBC \cite{b7}           & $+0.840$ & 0.0667 & $+0.900$ & 0.0833 & $+0.735$ & 0.1500 & $+0.900$ & 0.0833 \\
LogME \cite{b5}         & $+0.853$ & 0.0583 & $+0.900$ & 0.0833 & $+0.747$ & 0.1417 & $+0.900$ & 0.0833 \\
H-score \cite{b6}       & $+0.908$ & 0.0583 & $+0.800$ & 0.1333 & $+0.821$ & 0.1083 & $+0.800$ & 0.1333 \\
H-score shrunk \cite{b32} & $+0.934$ & 0.0417 & $+0.900$ & 0.0833 & $+0.850$ & 0.0583 & $+0.900$ & 0.0833 \\
TransRate \cite{b34}    & $+0.854$ & 0.0750 & $+0.900$ & 0.0833 & $+0.738$ & 0.1500 & $+0.900$ & 0.0833 \\
NLEEP \cite{b33}        & $+0.966$ & 0.0250 & $+0.900$ & 0.0833 & $+0.901$ & 0.0333 & $+0.900$ & 0.0833 \\
PACTran-G \cite{b35}    & $+0.889$ & 0.0833 & $+0.900$ & 0.0833 & $+0.846$ & 0.0750 & $+0.900$ & 0.0833 \\
SFDA \cite{b36}         & $+0.874$ & 0.0583 & $+0.900$ & 0.0833 & $+0.758$ & 0.1250 & $+0.900$ & 0.0833 \\
\hline
\end{tabular}}
\end{table}

\textbf{The source dataset.} The source relationship runs the other way, and no
published estimator addresses it. Table~\ref{tab:source} relates each dataset's
separability to its mean outgoing AUC. From this table, we observe that sources whose classes overlap more
heavily are associated with stronger transfer ($r = -0.922$, $p = 0.017$ for
the Silhouette Score; $r = -0.919$, $p = 0.017$ for the Fisher Ratio). Under
adaptation this is the only relationship in the study to reach the 0.05 level
on both descriptors. It is weaker and does not reach that level under direct
transfer ($r = -0.706$, $p = 0.133$), where it remains a consistent tendency in
sign.

\begin{table}[!t]
\caption{Source Structure versus Mean Outgoing AUC ($n=5$; Exact Permutation
$p$). No Reference Estimator Describes the Source Role.}
\label{tab:source}
\centering
\setlength{\tabcolsep}{3pt}
\resizebox{\columnwidth}{!}{%
\begin{tabular}{lcccc|cccc}
\hline
& \multicolumn{4}{c|}{DANN} & \multicolumn{4}{c}{Direct transfer} \\
Measure & $r$ & $p$ & $\rho$ & $p_\rho$ & $r$ & $p$ & $\rho$ & $p_\rho$ \\
\hline
Silhouette Score & $-0.922$ & 0.0167 & $-0.900$ & 0.0833 & $-0.706$ & 0.1333 & $-0.600$ & 0.3500 \\
Fisher Ratio     & $-0.919$ & 0.0167 & $-0.900$ & 0.0833 & $-0.706$ & 0.1250 & $-0.600$ & 0.3500 \\
\hline
\end{tabular}}
\end{table}

\textbf{Both endpoints.} Because the two roles are associated with separability
in opposite senses, a dataset that is a favorable target tends to be a weaker source.
This opposition is the structural basis of transfer asymmetry.
Table~\ref{tab:pairlevel} reports two pair-level analyses. The signed
difference in separability tracks the signed difference in AUC between the two
directions of a pair closely, in the same sense under both regimes and for both
descriptors ($r = -0.932$, $p = 0.017$ under adaptation): transfer favors the
direction running from the more overlapped domain toward the more separable
one. The absolute size of the difference, by contrast, carries no signal under
either regime (all $|r| \leq 0.33$, all $p \geq 0.28$). Two structurally
distant domains do not transfer worse than two structurally close ones.
The relevant quantity is which of the two domains is more separable, 
and not the magnitude of the difference.

\begin{table}[!t]
\caption{Pair-Level Analyses ($n=10$ Unordered Pairs; Dataset-Label Permutation
$p$). The Signed Difference Is $s(A)-s(B)$ against
$\mathrm{AUC}(A\!\rightarrow\!B)-\mathrm{AUC}(B\!\rightarrow\!A)$.}
\label{tab:pairlevel}
\centering
\setlength{\tabcolsep}{3pt}
\resizebox{\columnwidth}{!}{%
\begin{tabular}{llcc|cc}
\hline
& & \multicolumn{2}{c|}{DANN} & \multicolumn{2}{c}{Direct transfer} \\
Descriptor & Analysis & $r$ & $p$ & $r$ & $p$ \\
\hline
Silhouette & Signed difference vs signed $\Delta$AUC & $-0.932$ & 0.0167 & $-0.859$ & 0.0250 \\
           & Absolute gap vs mean pair AUC           & $-0.301$ & 0.3083 & $-0.321$ & 0.2833 \\
Fisher     & Signed difference vs signed $\Delta$AUC & $-0.932$ & 0.0167 & $-0.860$ & 0.0250 \\
           & Absolute gap vs mean pair AUC           & $-0.269$ & 0.3667 & $-0.294$ & 0.3167 \\
\hline
\end{tabular}}
\end{table}

\begin{figure}[!t]
\centerline{\includegraphics[width=\columnwidth]{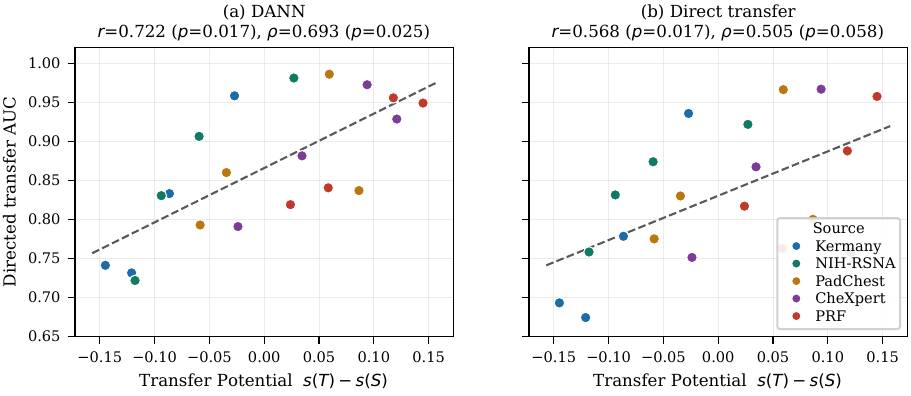}}
\caption{Transfer Potential versus directed transfer AUC across all twenty
ordered pairs, under (a) DANN and (b) direct transfer. The dashed line is the
linear trend and point color denotes the source dataset. A positive Transfer
Potential corresponds to a pairing in which the target is more separable than
the source.}
\label{fig:tp}
\end{figure}

Finally, Table~\ref{tab:tp} relates the Transfer Potential of
\eqref{eq:transferpotential} to directed AUC across all twenty ordered pairs,
and Fig.~\ref{fig:tp} shows the relationship. As a single interpretable
quantity combining both endpoints it is positively associated with transfer
performance in both regimes, accounting for about half the variance under
adaptation ($R^2 = 0.521$).

These relationships are not driven by any one dataset. Omitting each dataset in
turn leaves every correlation with its sign unchanged, and under adaptation the
target, source and pair-level associations retain magnitudes of at least 0.89,
0.91 and 0.78 respectively in the worst case.

\begin{table}[!t]
\caption{Transfer Potential versus Directed AUC ($n=20$ Directed Pairs;
Dataset-Label Permutation $p$)}
\label{tab:tp}
\centering
\setlength{\tabcolsep}{4pt}
\resizebox{\columnwidth}{!}{%
\begin{tabular}{llccccc}
\hline
Descriptor & Regime & $r$ & $p$ & $\rho$ & $p_\rho$ & $R^2$ \\
\hline
Silhouette & DANN            & $+0.722$ & 0.0167 & $+0.693$ & 0.0250 & 0.521 \\
           & Direct transfer & $+0.568$ & 0.0167 & $+0.505$ & 0.0583 & 0.323 \\
Fisher     & DANN            & $+0.720$ & 0.0167 & $+0.693$ & 0.0250 & 0.519 \\
           & Direct transfer & $+0.569$ & 0.0167 & $+0.505$ & 0.0583 & 0.324 \\
\hline
\end{tabular}}
\end{table}

\subsection{Predicting the Direction of Transfer}
\label{sec:direction}

The preceding relationship can be stated as a decision rule. For each of the
ten unordered pairs, the direction predicted to transfer better is the one with
positive Transfer Potential, that is, the direction running toward the more
separable domain. Table~\ref{tab:direction} reports how often this rule is
correct.

From Table~\ref{tab:direction}, we observe that under adaptation 
the rule identifies the better direction in all ten pairs
(exact binomial against chance, $p = 0.001$). Under direct transfer it is
correct in eight of ten ($p = 0.055$), the two exceptions being Kermany versus
NIH-RSNA and PRF versus PadChest. Substituting the Fisher Ratio reproduces both
counts exactly, as do GBC and the reference estimators, since all induce the
same ordering. The rule is also retained under dataset removal. Omitting each dataset in
turn, it is correct in all six remaining pairs on every one of the five
subsets.

We deliberately do not introduce an abstention band on small differences. With
ten pairs, any threshold wide enough to exclude the two direct-transfer
exceptions also discards six of the ten pairs, and the only basis for choosing
it would be that it removes the observed errors. We thus report the
unconditioned counts.

\begin{table}[!t]
\caption{Direction Prediction. Fraction of the Ten Unordered Pairs for Which
the Sign of the Transfer Potential Identifies the Better of the Two
Directions.}
\label{tab:direction}
\centering
\setlength{\tabcolsep}{5pt}
\begin{tabular}{llcc}
\hline
Descriptor & Regime & Correct & Binomial $p$ \\
\hline
Silhouette & DANN            & 10/10 & 0.0010 \\
           & Direct transfer & 8/10  & 0.0547 \\
Fisher     & DANN            & 10/10 & 0.0010 \\
           & Direct transfer & 8/10  & 0.0547 \\
\hline
\end{tabular}
\end{table}

\subsection{Robustness to the Feature Extractor}
\label{sec:robust}

The primary extractor was trained on NIH ChestX-ray14, the source of one of the
five evaluated datasets, so the entire analysis was repeated with a
DenseNet-121 trained on CheXpert, the source of another. The two networks order
the five datasets almost identically (Spearman $\rho = 0.90$ for the Silhouette
Score, $1.00$ for GBC). From Table~\ref{tab:robust}, we observe that all eight
relationships retain their sign under both descriptors, sixteen of sixteen in
total. Neither extractor assigns an inflated separability to its own training domain.
NIH-RSNA ranks second on separability under the NIH-trained weights but first under the
CheXpert-trained weights, and CheXpert ranks fourth under both. It should be noted that each extractor
overlaps one evaluated dataset, so this comparison demonstrates that the findings do
not depend on a single extractor-dataset overlap; it does not remove the
overlap.

\begin{table}[!t]
\caption{Robustness to the Feature Extractor: Pearson $r$ Under Two
Independently Trained DenseNet-121 Networks}
\label{tab:robust}
\centering
\setlength{\tabcolsep}{3pt}
\resizebox{\columnwidth}{!}{%
\begin{tabular}{llcc|cc}
\hline
& & \multicolumn{2}{c|}{Silhouette} & \multicolumn{2}{c}{Fisher Ratio} \\
Relationship & Regime & NIH & CheXpert & NIH & CheXpert \\
\hline
Separability $\rightarrow$ incoming AUC & DANN   & $+0.934$ & $+0.963$ & $+0.932$ & $+0.823$ \\
                                        & Direct & $+0.850$ & $+0.927$ & $+0.852$ & $+0.726$ \\
Separability $\rightarrow$ outgoing AUC & DANN   & $-0.922$ & $-0.761$ & $-0.919$ & $-0.909$ \\
                                        & Direct & $-0.706$ & $-0.572$ & $-0.706$ & $-0.842$ \\
Signed difference $\rightarrow$ $\Delta$AUC & DANN   & $-0.932$ & $-0.908$ & $-0.932$ & $-0.847$ \\
                                        & Direct & $-0.859$ & $-0.889$ & $-0.860$ & $-0.809$ \\
Transfer Potential $\rightarrow$ AUC    & DANN   & $+0.722$ & $+0.704$ & $+0.720$ & $+0.657$ \\
                                        & Direct & $+0.568$ & $+0.575$ & $+0.569$ & $+0.539$ \\
\hline
\end{tabular}}
\end{table}

\subsection{Class Balance}
\label{sec:balance}

Kermany and PRF are the only class-imbalanced datasets and they occupy the extremes of the separability scale, so the proposed descriptors were recomputed
on class-balanced subsamples of these two datasets, obtained by discarding
surplus images of the majority class ($1349$ and $1719$ images per class,
respectively), using the primary NIH-weighted extractor. The remaining three
datasets are already exactly balanced at $3000$ images per class, so balancing
leaves them unchanged and their values are carried over. Averaged over ten
random draws, Silhouette moves from $0.154$ to $0.137 \pm 0.002$ for Kermany
and from $0.009$ to $0.011 \pm 0.001$ for PRF, with Fisher Ratio tracking it
($0.349 \to 0.320 \pm 0.005$ and $0.021 \to 0.021 \pm 0.001$). Both shifts are
an order of magnitude smaller than the separability gap within which they occur.
Balanced Kermany remains more than twelve times as separable as balanced PRF,
and the full five-dataset ordering is preserved. A size-matched control drawn
at the original class ratio confirms that what little movement occurs is
attributable to balance rather than to the smaller sample, the balance
component exceeding the size component threefold for Kermany Silhouette
($-0.013$ versus $-0.004$).

Repeating every analysis on the resulting descriptor vector leaves all
conclusions intact. Under DANN, target structure predicts incoming AUC at
$r = +0.959$ ($p = 0.025$), source structure predicts outgoing AUC at
$r = -0.892$ ($p = 0.033$), Transfer Potential predicts directed AUC at
$r = +0.730$ ($p = 0.017$), and the sign of the performance asymmetry is
recovered for all ten dataset pairs ($p = 0.001$); the pair-level absolute gap
remains the one uninformative quantity. Every correlation retains its sign and
significance across all ten draws, and the direction rule is correct for all
ten pairs in each. Class imbalance thus does not generate the structural
differences the proposed descriptor measures, nor the relationships built on them.

%%%%%%%%%%%%%%%%%%%%%%%%%%%%%%%%%%%%%%%%%%%%%%%%%%%%%%%%%%%%%%%%%%%%%%%%%%%%%%%
\section{Discussion}
\label{sec:discussion}

Adaptation performance for pneumonia detection across these five chest
radiograph domains depends on the class structure of both endpoints, and the
two dependencies have opposite sign. Each is discussed in turn,
followed by the directional relationship that emerges from their combination.

\begin{figure}[!t]
\centerline{\includegraphics[width=\columnwidth]{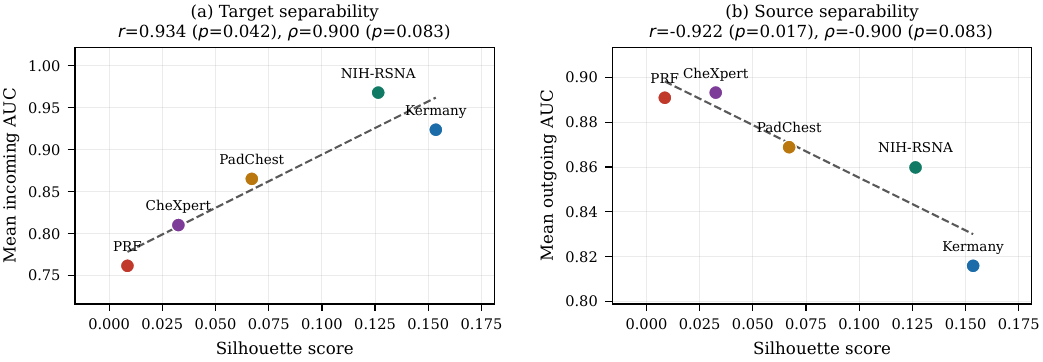}}
\caption{The two endpoints play opposite structural roles under adaptation.
(a) Targets with higher class separability receive higher mean incoming AUC.
(b) Sources with higher class separability produce lower mean outgoing AUC, so
more overlapped sources tend to transfer more strongly.}
\label{fig:endpoints}
\end{figure}

\subsection{Target Separability as a Reference Point}
\label{sec:disc_target}

Targets with better-separated classes are easier to adapt into
(Fig.~\ref{fig:endpoints}(a)), consistent with the cluster assumption: a target
whose classes are already distinct provides a clearer geometry onto which a
transferred decision boundary can align. We claim no novelty here. Every one of
the seven published estimators computed on the same representation reproduces
this relationship, and two of them match or exceed the proposed
descriptor (Table~\ref{tab:target}). The comparison establishes the target side
as a solved reference point and isolates the two aspects that remain unsolved,
namely the source and the direction.

\subsection{Source Overlap and Source Strength}
\label{sec:disc_source}

The source relationship is less expected, and no published
estimator describes it. Datasets whose classes overlap more heavily act as
stronger sources (Fig.~\ref{fig:endpoints}(b)), against the intuition that a
cleanly separable source, being an easier problem, should transfer best.

\begin{figure}[!t]
\centerline{\includegraphics[width=\columnwidth]{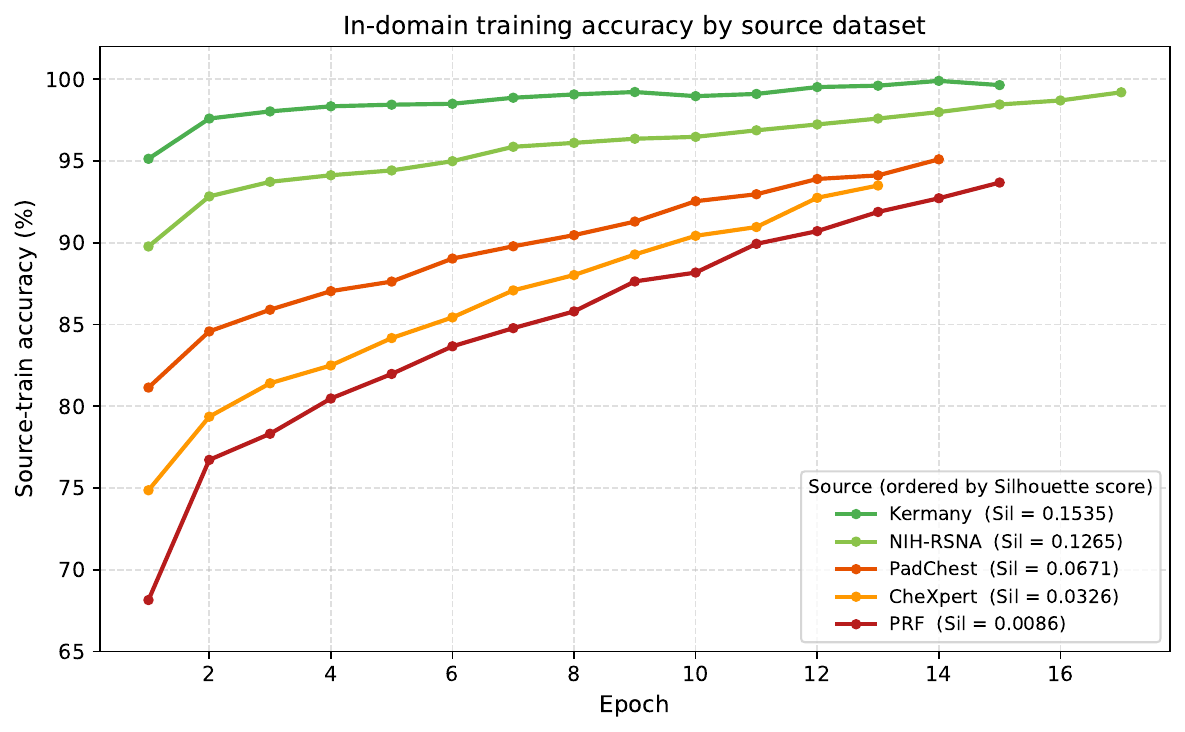}}
\caption{In-domain training accuracy of each dataset utilized as a source, ordered
by Silhouette Score. The most separable sources begin near their ceiling and
saturate within a few epochs; the more overlapped sources begin far lower and
improve gradually.}
\label{fig:traindyn}
\end{figure}

The training behavior of the source models is consistent with an explanation
in terms of the representations they learn. Fig.~\ref{fig:traindyn} shows
in-domain training accuracy for each dataset utilized as a source, ordered by
separability. From this figure, we observe that the two most separable sources begin between approximately 90\%
and 95\% accuracy and saturate within a few epochs, whereas the more overlapped
sources begin far lower and improve gradually. A source with tightly separated
classes allows the model to fit its task quickly, which suggests it can do so using
a narrow set of dataset-specific features that need not generalize. A source
with overlapping classes does not permit such a solution. Separating its classes at
all requires a broader, more distributed set of features, and those appear to
transfer better. The class overlap that makes a dataset more difficult in isolation
may thus also be the property that makes it a more effective source.

This is reported as a consistent observation rather than an established
mechanism. The direction holds under both regimes, under both descriptors and
under both feature extractors, and the training dynamics align with the
proposed account. It should be noted, however, that the dynamics are consistent with that account rather than
proof of it, since the rate at which a source is fit also depends on dataset
size and class balance. Confirmation requires a larger pool of domains.

\subsection{Both Endpoints and the Asymmetry of Transfer}
\label{sec:disc_asym}

Source and target occupy opposite structural roles: a dataset with
well-separated classes is a favorable target but a weaker source, and one with
heavy overlap the reverse, as illustrated by the inversion of Kermany and PRF.
Transfer asymmetry provides the clearest evidence that both endpoints contribute.
Adapting from A to B generally differs from B to A, and the preferred direction follows
the signed difference in separability, running from the more overlapped domain toward
the more separable one.

A target-only account cannot express this. Because it scores only the target
domain, it ignores which domain supplies the labels. Predictable transfer
asymmetry is thus direct evidence that the source domain's own structure
carries independent information. The present work departs from the estimators of
Section~\ref{sec:transferability}, which treat the source solely as a frozen
feature extractor and thus yield a property of the target alone. We verified
this empirically by computing seven of them on the same representation.

\subsection{Direction Rather Than Magnitude}
\label{sec:disc_direction}

The absolute structural gap between two domains showed no detectable
relationship with performance under either regime or either descriptor:
structurally distant domains do not transfer worse than close ones.
Transfer is governed by which of the two domains is more separable 
rather than by the extent of their difference. The relevant quantity
is thus a signed difference and not a distance. The Transfer Potential 
captures exactly this, and underlies the direction rule of Section~\ref{sec:direction}.

The limits of the proposed index should be noted. Transfer Potential does not predict performance more
accurately than target separability alone; the target term carries most of the
weight and the source term a smaller, consistent component. Its value lies in
interpretability, in expressing as one quantity that both endpoints matter and
enter with opposite sign.

\subsection{Consistency Across Regimes, Representations and Measures}
\label{sec:disc_regimes}

Every relationship reported here appears with the same sign under direct
transfer as under adaptation, under two independently trained feature
extractors, and under two separability measures built on different principles.
If these relationships had appeared only under adaptation,
they could plausibly be artifacts of adversarial training.
Because they appear equally without it, they are better understood
as properties of the domains that adaptation reflects rather than produces.
The relationships are weaker under direct transfer, and the
source-side effect in particular does not reach significance there, so that
regime corroborates the direction of the findings without independently
establishing them.

%%%%%%%%%%%%%%%%%%%%%%%%%%%%%%%%%%%%%%%%%%%%%%%%%%%%%%%%%%%%%%%%%%%%%%%%%%%%%%%

\subsection{Practical Implications}
\label{sec:disc_practical}

Characterizing the five datasets required one pass over each; evaluating the
same twenty directed pairings by training would have required twenty adaptation
runs, and the gap widens quadratically with the number of candidate domains.

The practical value of this result depends on which labels are available. The source-side result
needs only source labels, so a team holding several candidate public datasets
can rank them as sources before touching the target, preferring the more
overlapped one. The target-side and directional results need a labeled target
subset, which unsupervised adaptation does not assume; they apply where a small
validation set already exists, as it often does before clinical deployment, and
not to the fully unlabeled case. How small that subset may be is not settled
here. In neither case does the descriptor replace running adaptation: it
accounts for part of the variation in transfer performance, and is better utilized
to order candidates and set expectations than to select among them
automatically.

%%%%%%%%%%%%%%%%%%%%%%%%%%%%%%%%%%%%%%%%%%%%%%%%%%%%%%%%%%%%%%%%%%%%%%%%%%%%%%%

\subsection{Limitations}
\label{sec:disc_limits}

The analysis rests on five datasets, and as set out in Section~\ref{sec:stats}
no result at this scale could reach significance after correction for
multiplicity; the findings are hypothesis-generating, and the source-side
relationship in particular needs further domains. The scope is also narrow: 
the task is pneumonia detection from chest radiographs, with nothing here
establishing that the same relationships hold for other pathologies,
multi-class problems or modalities. Only one adaptation algorithm was examined,
so the claim that the relationships are not method-specific rests on DANN
together with the no-adaptation baseline.

Three measurement caveats are relevant to interpretation. The reported adapted AUC is
the best value attained on the target evaluation portion over training,
measuring what adaptation can reach rather than what a deployed model would
obtain. Both feature extractors were trained on one of the five evaluated
datasets, so neither is fully independent of the pool, although their agreement
indicates the conclusions do not hinge on a single overlap. The reference
estimators serve as target-side checks only, and LEEP and the Dirichlet variant
of PACTran could not be included at all (Section~\ref{sec:gbc}). Extending the
characterization beyond class separability in one learned feature space remains
an open direction.

%%%%%%%%%%%%%%%%%%%%%%%%%%%%%%%%%%%%%%%%%%%%%%%%%%%%%%%%%%%%%%%%%%%%%%%%%%%%%%%
\section{Conclusion}
\label{sec:conclusion}

Five chest radiograph domains were characterized by the class separability of
their features, obtained from one pass of a frozen DenseNet-121, and that
characterization was related to radiographic pneumonia detection performance across
all twenty directed source-target pairs under unsupervised domain adaptation.
Performance depends on both endpoints in opposite directions: more separable
targets received higher incoming AUC ($r = 0.934$, $p = 0.042$), while more
separable sources produced lower outgoing AUC ($r = -0.922$, $p = 0.017$). The
signed difference between the two identified the better direction of transfer
in 10 of 10 pairs ($p = 0.001$), whereas its magnitude carried no signal.
Every relationship held in sign under direct transfer, under a second feature
extractor and under a second separability measure.

The target-side relationship was reproduced by all seven published
transferability estimators computed on the same representation, and is treated
here as a reference point rather than a contribution. Those estimators do not
describe the source, and the opposition between the two endpoints is why the
direction of transfer can be anticipated at all. With five domains the finding
remains exploratory; confirming it at larger scale, across further adaptation
methods and beyond radiographic pneumonia detection, would establish domain
characterization as a practical step for teams deploying these models where the
cost of a misdirected adaptation run is highest.

\section*{Data Availability}
The Kermany, CheXpert, NIH ChestX-ray14, RSNA and PadChest datasets are
publicly available from their original distributors. The PRF dataset is a
private collection and is not publicly available. Code, pre-computed structural
descriptors and the full statistical pipeline are available at
\url{https://github.com/mHealthBuet/CXR-Domain-Structure-UDA}.

\section*{Conflict of Interest}
The authors declare no conflict of interest.

%%%%%%%%%%%%%%%%%%%%%%%%%%%%%%%%%%%%%%%%%%%%%%%%%%%%%%%%%%%%%%%%%%%%%%%%%%%%%%%

\end{document}